\documentclass[apj]{emulateapj}

\usepackage{amsmath}

\shorttitle{Multi-Planet Gap Depths}
\shortauthors{Duffell \& Dong}

\begin{document}

\author{Paul C.~Duffell\altaffilmark{1} and Ruobing Dong\altaffilmark{1,2}}
\altaffiltext{1}{Astronomy Department and Theoretical Astrophysics Center, University of California, Berkeley, CA 94720}
\altaffiltext{2}{Hubble Fellow}
\email{duffell@berkeley.edu}

\title{Shallow Cavities in Multiple-Planet Systems}

\begin{abstract}

Large cavities are often observed in protoplanetary disks, which might suggest the presence of planets opening gaps in the disk.  Multiple planets are necessary to produce  a wide cavity in the gas.  However, multiple planets may also be a burden to the carving out of very deep gaps.  When additional planets are added to the system, the time-dependent perturbations from these additional satellites can stir up gas in the gap, suppressing cavity opening.  In this study, we perform two-dimensional numerical hydro calculations of gap opening for single and multiple planets, showing the effect that additional planets have on the gap depths.  We show that multiple planets produce much shallower cavities than single planets, so that more massive planets are needed in the multiple-planet case to produce an equivalent gap depth as in the single-planet case.  To deplete a gap by a factor of $100$ for the parameters chosen in this study, one only requires $M_p \approx 3.5 M_J$ in the single-planet case, but much more massive planets, $M_p \approx 7 M_J$ are required in the multiple-planet case.  This requirement of high-mass planets implies that such planets may be detectable in the next generation of direct imaging projects, in gaps whose depths are constrained to be sufficiently deep by ALMA.

\end{abstract}

\keywords{hydrodynamics --- planetary systems: protoplanetary disks}

\section{Introduction}
\label{sec:intro}

Many observed protostellar disks exhibit features broadly consistent with the presence of planets.  In particular, transitional disks can have cavities or gaps in their disk structure, and it is suspected in these cases that this could be due to a planet or several planets which have cleared their orbits of dust and gas \citep[see a recent review by][]{espaillat14}.

Spectral energy distributions of these disks can exhibit deficits, particularly around $\sim1-10 \mu m$.  This suggests that the warm dust in the inner part of these disks is somewhat depleted \citep{calvet05, espaillat07, espaillat10}. Cavities have also been observationally resolved, in both near infrared (NIR) by direct imaging of polarized scattered light \citep[e.g.][]{canovas13, quanz13, garufi13, avenhaus14, mayama12, hashimoto12}, and at mm wavelengths by observing the dust continuum and/or molecular line emissions \citep[e.g.][]{hughes09, andrews11, bruderer14, casassus13, vandermarel13, zhang14, perez14, perez14hd142527}. Resolved observations at these two spectral windows probe the spatial distribution of different disk components, which is important because different disk components do not always have the same spatial distribution \citep{dong12cavity}. NIR images trace the disk surface structure in the small grains ($\lesssim\micron$) as disks are generally optically thick at NIR, while optically thin mm continuum or molecular line emissions can be used to constrain the surface density of the gas and the big grains ($\sim$mm). 

It is very often non-trivial to ``translate'' observed image morphology at NIR and visibility measurements at mm to underlying physical density structures of various disk components, and radiative transfer simulations are often needed to make the connection.  Ideally one would like to constrain the depletion factor of the gas $\delta_{\rm gas} = \Sigma_{\rm edge} / \Sigma_{\rm gap}$ (where $\Sigma$ is the gas surface density), but NIR scattered light and millimeter continuum emission only probe the distribution of dust grains.  Therefore $\delta_{\rm gas}$ is typically inferred in one of two ways: (1) optically thin molecular line emission data can be fit to disk models to directly pin down $\delta_{\rm gas}$, or (2) the cavity depletion factor of the small grains $\delta_{\rm small}$ can be measured, and the gas depletion can then be constrained based on the assumption $\delta_{\rm gas}=\delta_{\rm small}$, as small grains are generally considered to be well mixed with the gas due to their small stopping time.

Both methods to infer $\delta_{\rm gas}$ are very challenging, because they necessitate both high resolution and high sensitivity to measure the cavity.  Only a few cavities in transitional disks have been modeled and measured in detail.  For example, using rotational transitions of various CO isotopes, \citet{bruderer14} found that observations are consistent with a model in which $\Sigma_{\rm gas}$ drops by a factor of $\sim~$12 inside the $\sim~$60 AU gap in IRS~48, and by at least a factor of 110 inside 20~AU.  \citet{perez14hd142527} found that depleting gas by a factor of $\sim~$50 inside the cavity in HD~142527 can fit their data.  Constraints on $\delta_{\rm gas}$ from these type of observations often depend on the assumptions made in the models, such as a certain undepleted/fiducial surface density profile and simplified disk model (which is generally assumed to be axisymmetric and to have a very sharp cavity edge), and are also limited by the quality of current data.  These constraints are generally only good to a factor of a few. By fitting both $H$~band images and the SED of PDS~70 simultaneously, \citet{dong12pds70} determined $\delta_{\rm small}\sim1000$ in the cavity of PDS~70, assuming a uniform depletion throughout the cavity. 

The ALMA project will improve the accuracy and precision of these numbers considerably, and will substantially increase the number of transitional disks whose cavities are measured and modeled.  We are therefore very close to observationally constraining the interaction between planets and their birth environments.  For the few examples studied so far, cavities are found to be depleted of gas by at least an order of magnitude, but possibly more.  This number (the depletion factor) should constrain the properties of the disks and the planets inhabiting these disks.  In order to make this connection, one must study the physical problem of how empty gaps can become as a function of disk and planet parameters.

Several studies have attempted to use semi-analytic theory to describe gap profiles \citep[e.g.][]{2004ApJ...612.1152V, 2006Icar..181..587C}.  These calculations have predicted highly depleted gaps (depletion factors are roughly exponential with planet mass) but numerical studies appear to be inconsistent with this result.

Recently, a few numerical studies have explicitly explored the question of how gap depth depends on planet and disk parameters.  \cite{2013ApJ...769...41D} calculated gap opening for low-mass planets in the weakly nonlinear regime, and found that for a planet-to-stellar mass ratio $q$ in a disk with viscosity $\alpha$ and aspect ratio $h/r$, the gap depth scales as 

\begin{equation}
\Sigma_{\rm gap} / \Sigma_0 = 0.09  \left({q \over 10^{-3}} \right)^{-2} \left( {\alpha \over .01} \right)^1 \left( { h/r \over 0.05 } \right)^5
\label{eqn:duff}
\end{equation}

so that gap depth exhibits power-law rather than exponential behavior (though the scaling with $h/r$ was not actually measured in that study).  More recently, \cite{2014ApJ...782...88F} explored the parameter space of giant planets, and found an empirical scaling relation for Jupiter-like planets:

\begin{equation}
\Sigma_{\rm gap} / \Sigma_0 = 0.14 \left({q \over 10^{-3}} \right)^{-2.1} \left( {\alpha \over .01} \right)^{1.4} \left( { h/r \over 0.05 } \right)^{6.5}
\label{eqn:fung}
\end{equation}

suggesting that \cite{2013ApJ...769...41D} captured a reasonably correct scaling with $q$ and $\alpha$ in the weakly nonlinear regime, but the scaling with $h/r$ may be stronger.  These studies suggest that deep cavities with depletion by a few orders of magnitude $\Sigma_{\rm gap} / \Sigma_0 \sim 10^{-2}$ are reasonable to expect for gas giants in typical disks.

However, all systematic quantitative numerical calculations of gap depths to date have focused on single-planet systems, though multiple-planet studies have begun to explore this question somewhat, see \cite{2011ApJ...729...47Z, 2011ApJ...738..131D, 2014arXiv1411.6063D}.  Gap opening in multiple-planet systems may be more complex, because the gap belonging to each planet is subject to periodic driving from neighboring planets, which may stir up the gas, having the overall effect of shallower gaps.  Such effects are also unlikely to be captured in current one-dimensional disk models, as the dynamics are complex and time-dependent.

In this letter, we investigate gap opening in multiple-planet systems, studying the specific question of whether the presence of additional planets in the system has a strong impact on gap depths.  We find that additional planets in the system indeed complicate the gap-opening process, resulting in much shallower cavities.  This means that producing deep cavities with multiple-planet systems may require very large planets, and this may be a problem if such large planets are not directly imaged.

\section{Numerical Setup}
\label{sec:numerics}

\begin{figure*}
\epsscale{1.0}
\plotone{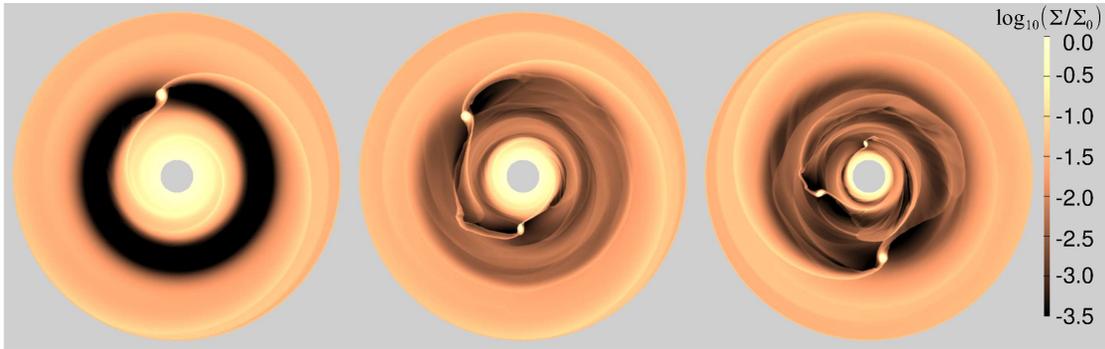}
\caption{ Logarithm of gas surface density demonstrating the significant difference in gap depths between the single and multiple-planet cases (Left panel is a single planet, and the center and right panels display two and three-planet systems, respectively).  Each planet in this figure has a mass of $4 M_J$.  A single planet of this mass carves out a gap of depth $\Sigma_{\rm gap} / \Sigma_{\rm edge} \sim 3 \times 10^{-3}$, whereas two or three planets of this mass produce a much shallower cavity, $\Sigma_{\rm gap} / \Sigma_{\rm edge} \sim 0.1$.  Additionally, multiple-planet systems generate a much more complex and dynamic density structure in the cavity, due to the time-dependent stirring of adjacent planet torques.
\label{fig:2d} }
\end{figure*}

\begin{figure}
\epsscale{1.0}
\plotone{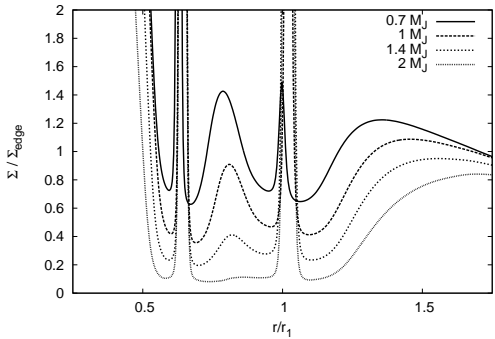}
\caption{ Averaged surface density (averaged both in azimuth and in time) in a two-planet system.  Each planet lives inside its own gap until the planets' combined Hill radii are of order the separation (Equation \ref{eqn:qcrit}).  For planets larger than this critical mass, the small gaps merge to produce a large gap.  However, when the planets are this large, they are capable of influencing the streamlines in each other's gaps, potentially complicating the gap-opening process.
\label{fig:merge} }
\end{figure}

This study performs a numerical integration of the two-dimensional isothermal fluid equations

\begin{eqnarray}
\partial_t \Sigma + \nabla \cdot ( \Sigma \vec v ) ~& = &~ 0 
\label{eqn:cons1}
\\
\partial_t ( \Sigma v_j ) + \partial_i ( \Sigma v_i v_j + P \delta_{ij} + \sigma_{ij} ) ~& = &~ -\Sigma \partial_i \phi ~~~~~
\label{eqn:cons2}
\end{eqnarray}

\begin{equation}
P = c^2 \Sigma,
\end{equation}

where $\Sigma$ is surface density, $\vec v$ is velocity, $P$ is pressure, $c$ is the sound speed, $\sigma$ is the viscous stress tensor, and $\phi$ is the gravitational potential of the central point mass and the perturbing planets.  In this study, the disk is assumed to be globally isothermal, so that the sound speed $c$ is a constant everywhere in the disk.  For the current study, the sound speed is chosen to give an aspect ratio of $h/r = 0.05$ in the vicinity of the outermost planet.

The numerical integration is carried out using the DISCO code, a moving-mesh hydrodynamics code tailored to the study of gaseous disks.  A resolution of 512 radial zones is employed, logarithmically spaced between $r_{\rm min} = 0.2 ~ r_1$ and $r_{\rm max} = 2.0 ~ r_1$, where $r_1$ is the radial position of the outermost planet.  This implies a resolution of $h/\Delta r = 11$ zones per scale height in this planet's vicinity.  \cite{2013ApJ...769...41D} argued that this resolution results in an effective numerical viscosity corresponding to $\alpha_{\rm num} \approx 2 \times 10^{-5}$, well below the chosen viscosity $\alpha = 10^{-2}$ in this study.

The initial disk profile is taken to be

\begin{equation}
\Sigma(r) = \Sigma_0 (r/r_0)^{-3/2}
\end{equation}

\begin{equation}
\Omega(r) = \sqrt{ GM_*/r^3 }
\end{equation}

\begin{equation}
v_r = -{3 \over 2} {\nu \over r}
\end{equation}

The viscosity in the disk is given an $\alpha$ prescription,

\begin{equation}
\nu = \alpha c h \propto r^{3/2}
\end{equation}

so that $\nu \Sigma =$ constant, and the background flow is a consistent inward accretion.

The potential from the protostar and its satellites is given by 

\begin{equation}
\phi(\vec x) = - G M_* \left( { 1 \over |\vec x| } + \sum_p {q \over \sqrt{(\vec x - \vec x_p)^2 + \epsilon^2}} \right),
\end{equation}

with smoothing length $\epsilon$ chosen in this study to be half of a scale height in the vicinity of each planet.  Each perturbing planet is given the same mass $M_p = q M_*$ so as to simplify the range of parameter space studied.  All planets are kept on fixed circular orbits, and the motion of the central protostar is neglected, though it is possible that this motion would also induce some gap stirring for large $q$.

The planets are placed in 2:1 resonance with one another, so that $r_2 = r_1 / 2^{2/3}$ and $r_3 = r_2 / 2^{2/3}$.

The planet-disk systems are run for thousands of orbits, until a steady-state is reached.  For the multiple-planet cases, the ``steady-state" is really quasi-steady, as the solution is highly time-dependent, but a relaxed time-averaged state is found consisting of a consistent inward flow through the disk.

\section{Results}
\label{sec:results}

Figure \ref{fig:2d} presents the most basic result of gap opening in multiple-planet systems.  For $4 M_J$ planets, the two and three-planet systems have much shallower gaps than the single-planet system.  The fact that the planets can interact with one another's gaps is related to the fact that the gaps have merged with one another.

For small enough planet mass ($M_p \lesssim M_J$), each gap is confined to the planet carving it out.  If the planet's mass is increased, the gap width grows until the two gaps merge.  This is shown in Figure \ref{fig:merge}, which plots azimuthally averaged surface density in the two-planet system for several planet masses.  The $M_p = 1.4~M_J$ system opens two distinguishable gaps, while the $M_p = 2~M_J$ system opens a single gap containing both planets.  This ``gap merging" should happen roughly when the combined Hill radii of the two planets are of order the separation between the planets:

\begin{equation}
r_{\rm H 1} + r_{\rm H 2} \sim \Delta r,
\end{equation}

where the Hill radius $r_H \approx r (q/3)^{1/3}$.  So, one should expect the gaps to merge when

\begin{equation}
{r_{\rm H 1} + r_{\rm H 2} \over \Delta r} = C,
\end{equation}

where $C$ is some order-unity constant.  In this study, the empirical result $C \approx 0.36$ is found, implying a critical mass ratio

\begin{equation}
q_{\rm crit} \approx 0.017 \left( {\Delta r \over \bar r} \right)^3,
\label{eqn:qcrit}
\end{equation}

where $\bar r$ is the mean of the two radii, $\bar r = (r_1 + r_2)/2$.  Though this scaling with separation is largely speculative, as the variation of gap properties with planet separation was not explicitly explored in this study.  Equation (\ref{eqn:qcrit}) also assumes both planets have the same mass.  Inserting the separation for a 2:1 resonance $\Delta r_{2:1} \approx 0.45 ~\bar r$ into this formula, this critical mass appears at the planet mass $M_p \sim 1.6 M_J$ for the separation in the current study.  On the other hand, this criterion also implies the planets are massive enough to gravitationally influence each other's gaps.  Therefore, as soon as the gaps are capable of merging, producing a large cavity, the planets are strong enough to stir up gas in each other's neighborhood.  In short, a wide cavity can only be produced at the cost of keeping the cavity shallow.  For very low viscosities or very thin disks, the gap might still be deep enough to represent a large depletion factor, but for the chosen disk parameters ($\alpha = 0.01$, $h/r = 0.05$) the gap is only depleted by about a factor of ten at $q \sim q_{\rm crit}$, which is shallower than depths inferred in some observations.

This is most clearly demonstrated in Figure \ref{fig:2d}, which shows single, double, and triple planet systems, for planets of mass $M_p = 4 M_J$.  The single planet easily carves out a deep gap ($\Sigma_{\rm gap}/\Sigma_{\rm edge} \sim 3 \times 10^{-3}$), but multiple-planet systems exhibit much shallower cavities ($\Sigma_{\rm gap}/\Sigma_{\rm edge} \sim 0.1$).

\begin{figure}
\epsscale{1.0}
\plotone{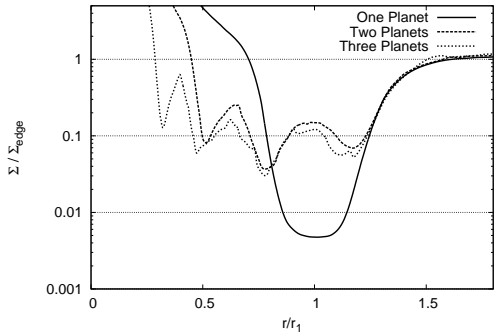}
\caption{ Azimuthally averaged version of Figure \ref{fig:2d} (gas near the planet is excluded from the azimuthal average).  Averaged surface density is plotted for planets with mass $= 4 M_J$, comparing a single planet with two and three planets (all with $m_p = 4 M_J$).
\label{fig:profile} }
\end{figure}

Figure \ref{fig:profile} is an azimuthally averaged version of Figure \ref{fig:2d} (in this average, gas within the planet's Hill sphere is ignored, as this depends sensitively on how gas behaves very close to the planet, which is not well-captured in this numerical calculation).  The cavity is of course wider when more planets are added, but clearly the cavity is also much more shallow, by more than an order of magnitude.  This may provide tension for observations of these cavities, if future observations can constrain gas depths to be as low as $\Sigma_{\rm gap}/\Sigma_{\rm edge} \sim 10^{-2}$ (current direct observations of the gas density cannot yet constrain the gap depth this precisely).  Deep cavities require much more massive planets than in the single-planet case.

\begin{figure}
\epsscale{1.0}
\plotone{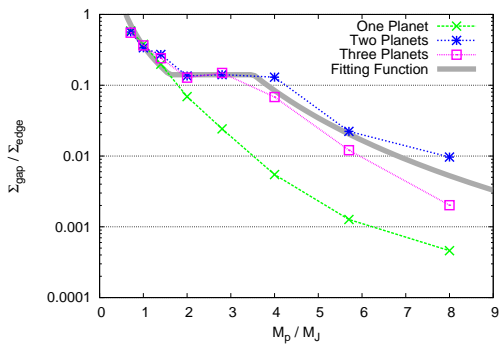}
\caption{ Gap depth as a function of planet mass, for one-planet, two-planet and three-planet systems.  The fitting function (\ref{fig:fitting}) is plotted alongside these data.
\label{fig:gapdepth} }
\end{figure}

Figure \ref{fig:gapdepth} shows how this result depends on planet mass.  Depletion is calculated by $\Sigma_{\rm gap}/\Sigma_{\rm edge}$, where $\Sigma_{\rm gap}$ is the value of the azimuthally averaged density at the outermost planet's orbital radius (again, ignoring material within the planet's Hill sphere), and $\Sigma_{edge}$ is the surface density at $r = 1.6 r_1$, which is roughly where $\Sigma(r)$ attains its maximum.

For values of $q$ below the critical threshold (\ref{eqn:qcrit}), the gap depths are reasonably consistent with the scalings (\ref{eqn:duff}) or (\ref{eqn:fung}).  At $q = q_{crit}$ ($\sim 1.6 M_J$ for the chosen parameter set), the cavity stops depleting in the multiple-planet cases, and $\Sigma(q)$ exhibits a plateau until about $4 M_J$, after which $\Sigma(q) \sim q^{-4}$.  

The value of the surface density in this plateau is probably the most important observable in this study.  A reasonable estimate for the surface density in this case appears to be given by $\Sigma_{\rm plateau} \sim \Sigma(q_{\rm crit})$, where $\Sigma(q)$ is given by either (\ref{eqn:duff}) or (\ref{eqn:fung}).  For very large-mass planets, $q > 2.2~q_{\rm crit}$, the plateau ends and a steep scaling of $\Sigma_{\rm gap}(q) \sim q^{-4}$ is found, though it should be stressed that this scaling is uncertain, as it is only based on a few data points.  All of this can be summarized by the following approximate fitting formula:

\begin{equation}
{\Sigma_{\rm gap} \over \Sigma_{\rm edge}} \approx \left\{ \begin{array}
				{l@{\quad \quad}l}
				s(q) & q < q_{\rm crit} 	\\  
    			s(q_{\rm crit}) & q_{\rm crit} < q < 2.2~q_{\rm crit} 	\\ 
    			s(q_{\rm crit})\left( {q \over 2.2~q_{\rm crit} } \right)^{-4} & q > 2.2~q_{\rm crit}				\\
    			\end{array} \right.    			
\label{fig:fitting}
\end{equation}

where

\begin{equation}
s(q) = 0.3  \left({q \over 10^{-3}} \right)^{-2},
\end{equation}

assuming the disk parameters chosen in this study, $\alpha = 0.01$, $h/r = 0.05$.  $q_{crit}$ is given by equation (\ref{eqn:qcrit}).  $s(q)$ is the surface density scaling for low-mass planets ($q \ll q_{\rm crit}$) and the transition at $2.2~q_{\rm crit}$ and the scaling $q^{-4}$ at high $q$ is simply a fit to the high-mass data.  It is unclear how the high-mass behavior should scale with disk parameters $\alpha$ or $h/r$; this could be checked in a more complete study.

\section{Discussion}
\label{sec:discussion}	

Deep gaps are challenging to produce with multiple-planet systems.  Very massive planets are necessary to explain density depletion factors of order $10^{-2}$, possibly $7$ Jupiter masses or larger, compared with the $\sim 3.5 M_J$ necessary in the single-planet case.  Once the necessary planet mass enters the detectable planet regime in observations, it may be possible to either confirm or rule out planets as the agents responsible for some large cavities.

The caveat to this is that low viscosities might result in deeper gaps, and low viscosity might be expected at large radii ($10 - 100$ AU) due to the fact that the disk is not significantly ionized at large distances from the protostar, meaning that the magnetorotational instability may not generate a significant turbulent disk viscosity at these radii \citep[e.g.][]{2006AnA...445..205I}.  However, since the gap stirring is produced dynamically, the effective viscosity induced may be independent of the disk viscosity and therefore the same might be true of the gap depth.  Gap depths are also known to be highly sensitive to $h/r$ in single-planet systems, as seen by the steep dependence in equations (\ref{eqn:duff}) and (\ref{eqn:fung}).  A more complete parameter survey is warranted.

A deep gap implies a very massive planet and therefore it is reasonable to ask whether resolution and sensitivity are high enough to detect such a massive planet.  Recently, several planet candidates have been found in transition disks, including T~Cha \citep{huelamo11}, LkCa~15 by \citep{kraus12}, HD~100546 by \citep{quanz13hd100546, brittain14}, HD~142527 \citep[see also \citealt{biller12}]{close14}, and HD~169142 by \citep{reggiani14}. The masses of the planet candidates can be as low as $\sim6M_{\rm J}$ (or even smaller, \citealt{kraus12}), although they are still very model dependent.  

Additionally, it may soon be possible to directly image the planet; two survey scale direct imaging projects each with dedicated instruments are currently being carried out: the Gemini Planet Imager Exoplanet Survey (GPIES) \citep{macintosh08}, and the Spectro-Polarimetric High-contrast Exoplanet REsearch (SHPERE) campaign with VLT \citep{beuzit08}. More results are expected in the next few years.

The next generation of observations of transition disks may therefore be able to detect the planet which is producing the gap.  If the planet is not detected, and if the gas in cavity can be constrained to a depth of $\Sigma_{\rm gap} / \Sigma_{\rm edge} \lesssim 10^{-2}$ or smaller, it should provide tension to the hypothesis that the cavity is produced by planets in the first place.

\acknowledgments

This work is partially supported by NASA through Hubble Fellowship grant HST-HF-51320.01-A awarded by the Space Telescope Science Institute, which is operated by the Association of Universities for Research in Astronomy, Inc., for NASA, under contract NAS 5-26555.  Resources supporting this work were provided by the NASA High-End Computing (HEC) Program through the NASA Advanced Supercomputing (NAS) Division at Ames Research Center.

We are grateful to Eugene Chiang and Ji-Ming Shi for helpful comments and discussions.  We would like to thank the anonymous referee for his or her thoughtful review.

\bibliographystyle{apj}

\end{document}